\documentclass[preprint,showpacs,preprintnumbers,amsmath,amssymb]{revtex4}
\usepackage{graphicx}
\usepackage{dcolumn}
\renewcommand{\b}[1]{\mbox{\boldmath $#1$}}

\begin{document}
\title{Surface gravity waves in deep fluid at vertical shear flows}

\author{G.~Gogoberidze}
\email{gogober@geo.net.ge}
\author{L.~Samushia}
\author{G.~D.~Chagelishvili}%
\author{J.~G.~Lominadze}%
\affiliation{Center for Plasma Astrophysics, Abastumani
Astrophysical Observatory, Ave. A. Kazbegi 2a, Tbilisi 0160,
Georgia}
\author{W.~Horton}%
\affiliation{Institute for Fusion Studies, The University of Texas
at Austin, Austin, Texas 78712}

\date{\today}

\begin{abstract}

Special features of surface gravity waves in deep fluid flow with
constant vertical shear of velocity is studied. It is found that
the mean flow velocity shear leads to non-trivial modification of
surface gravity wave modes dispersive characteristics. Moreover,
the shear induces generation of surface gravity waves by internal
vortex mode perturbations. The performed analytical and numerical
study provides, that surface gravity waves are effectively
generated by the internal perturbations at high shear rates. The
generation is different for the waves propagating in the different
directions.  Generation of surface gravity waves propagating along
the main flow considerably exceeds the generation of surface
gravity waves in the opposite direction for relatively small shear
rates, whereas the later wave is generated more effectively for
the high shear rates. From the mathematical point of view the wave
generation is caused by non self-adjointness of the linear
operators that describe the shear flow.

\end{abstract}

\maketitle

\section{Introduction}               

Generation of surface gravity waves (SGW), that are the best known
sea and oceanic waves, are naturally associated with winds.
Momentum transfer from wind to undulating movement of the ocean
which is the basic mechanism of surface waves generation, is
investigated since Kelvin's pioneering work~\cite{K71}.
Independent and inter-complementary theories of Fillips~\cite{P57}
and Miles~\cite{M57,M59,M61,M62} provide the basics of theoretical
understanding of surface wave generation by wind. Fillips'
resonant mechanism is responsible for  excitation and initial
rising of wave motion on unexcited surface of the fluid; Miles'
mechanism --- energy transfer from wind to fluid as a consequence
of wind shear flow and surface waves interaction --- is
responsible for subsequent amplification of the waves. According
to Miles' mechanism the energy source is wind shear flows situated
outside the fluid. Other ways of SGW generation have been also
studied such as the possibility of SGW generation by
earthquakes~\cite{K62,K72} and theory of SGW generation by
intra-fluid explosions~\cite{KK59}. In theories mentioned above
sources of SGW generation are extrinsic for the fluid.

The question arises as to whether intrinsic for the fluid sources
(shear flows and vortex perturbations for example) could generate
SGW.

This question become especially interesting in view of the
impressive progress made in the understanding of spectrally stable
shear flow phenomena by hydrodynamic community in the past ten
years. The early transient period for the perturbations has been
shown to reveal rich and complicate behavior in smooth (without
inflection point) shear flows. Particularly, it has been shown
that the linear dynamics of perturbations in the flows are
accompanied by intense temporal energy exchange processes between
background flow and perturbations and/or between different modes
of perturbations. From the mathematical point of view these
effects are caused by the non-self adjointness of the linear
operators in shear flows and are adequately described in the
framework of so-called non-modal approach (see,
e.g.,~\cite{GL65,CD90,CRT96}). The non-modal approach implies a
change of independent variables from the laboratory frame to a
moving frame and the study of temporal evolution of {\it spatial
Fourier harmonics} (SFHs) of perturbations without any spectral
expansion in time.

We examine the linear dynamics of surface waves and internal
perturbations in deep fluid in the absence of wind and presence of
the fluid flow with vertical shear of velocity. Dispersive
characteristics of shear modified SGWs as well as linear mechanism
of surface waves generation in deep fluid by internal
perturbations is studied in detail in the framework of the
non-modal approach.

The paper is organized as follows: mathematical formalism is
presented in Sec. II. Shear modified SGWs and their generation is
analyzed in  Sec. III. Applications of the obtained results to the
concrete physical problems are discussed in Sec. IV. Conclusions are
given in Sec. V.

\section{Mathematical Formalism}

Consider deep fluid with flat outer surface at $z=0$ and constant
shear flow $\b U_0=(A z,0,0)$ for $z<0$. The shear parameter $A$
is considered to be positive for the simplicity. The gravitational
field is considered to be uniform $\b g = (0,0,-g)$. Generally,
four modes of perturbation: SGW, internal gravity waves, sound
waves and vortex mode can exist in the system. To reduce the
mathematical complications as much as possible but still keep the
basic physics of our analysis, we consider fluid to be
incompressible (which drops sound waves) and neglect the
stratification effects (assuming that frequency of internal
gravity waves is much less in comparison with the frequency of
SGWs, i.e., considering internal gravity waves as aperiodic/vortex
mode perturbations). In further analysis we ignore also effects of
viscosity. After these simplifications we maintain just two modes
of perturbation - SGW and vortex mode and write up the following
differential equations for the linear dynamics of perturbations of
velocity ($\b u^\prime$) and normalized pressure
($p^\prime=p/\rho_0$):

\begin{equation}
\frac{\partial u^\prime_x}{\partial x} + \frac{\partial
u^\prime_y}{\partial y} + \frac{\partial u^\prime_z}{\partial z}
 = 0,\label{eq:p1}
\end{equation}
\begin{equation}
\frac{\partial u^\prime_x}{\partial t}+A z \frac{\partial
u^\prime_x}{\partial x } + A u^\prime_z=-\frac{\partial
p^\prime}{\partial x},\label{eq:p2}
\end{equation}
\begin{equation}
\frac{\partial u^\prime_y}{\partial t}+A z\frac{\partial
u^\prime_y}{\partial x}= -\frac{\partial p^\prime}{\partial y},
\label{eq:p3}
\end{equation}
\begin{equation}
\frac{\partial u^\prime_z}{\partial t}+A z \frac{\partial
u^\prime_z}{\partial x} =-\frac{\partial p^\prime}{\partial z},
\label{eq:p4}
\end{equation}
with the boundary condition on the surface $z=0$:
\begin{equation}
\left(\frac{\partial p^\prime}{\partial t} - g u^\prime_z
\right)\bigg|_{z=0}=0. \label{eq:b1}
\end{equation}

Further we use the standard technique of the non-modal approach
\cite{GL65}: introduction of co-moving variables $(x^\prime=x+ A z
t, y^\prime=y, z^\prime=z, t^\prime=t)$ let us to transform
spatial inhomogeneity presented in
Eqs.~(\ref{eq:p1})-(\ref{eq:b1}) to temporal one. Then, after the
Fourier transform with respect to $x^\prime$ and $y^\prime$:
\begin{equation}
\b u^\prime(\b r,t)=\frac{1}{4\pi^2} \int\b u(k_x, k_y,z',t)
\exp\left[i( k_x x' + k_y y')\right] d k_x d k_y, \label{eq:fou}
\end{equation}
the dynamic equations are reduced to:
\begin{equation}
 i k_x u_x + i k_y u_y +\left(\frac{\partial}{\partial z^\prime} -
 i A t^\prime k_x\right) u_z = 0, \label{eq:p5}
\end{equation}
\begin{equation}
\frac{\partial u_x}{\partial t^\prime} + A u_z = - i k_x p,
\label{eq:p6}
\end{equation}
\begin{equation}
\frac{\partial u_y}{\partial t^\prime} = -i k_y p,
 \label{eq:p7}
\end{equation}
\begin{equation}
\frac{\partial u_z}{\partial t^\prime} =
-\left(\frac{\partial}{\partial z^\prime}-i A t^\prime
k_x\right)p,
 \label{eq:p8}
\end{equation}
\begin{equation}
\left(\frac{\partial p}{\partial t^\prime} - g u_z
\right)\bigg|_{z^\prime=0}=0. \label{eq:b2}
\end{equation}
Hereafter primes of $z^\prime$ and $t^\prime$ variables will be
omitted.

From this set we readily obtain the following equation for the
perturbation of vertical component of velocity:
\begin{equation}
\frac{\partial}{\partial t} \left(\left[\widetilde k^2 - \left(
\frac{\partial}{\partial z} -i A t k_x \right)^2 \right] u_z
\right) = 0, \label{eq:pr2}
\end{equation}
where $\widetilde k = \sqrt {k_x^2 + k_y^2}$.

All other perturbed quantities ($u_x,~u_y$ and $p$) can be readily
expressed through $u_z$ combining
Eqs.~(\ref{eq:p5})-(\ref{eq:p8}); e.g., for $p$ we have:
\begin{equation}
p = -\frac {1}{\widetilde k^2} \left(\frac{\partial}{\partial t}
\left[ \left( \frac{\partial}{\partial z} - i A t k_x \right) u_z
\right] - i A k_x u_z \right), \label{eq:pr1}
\end{equation}

Integration of Eq.~(\ref{eq:pr2}) with respect to time yields:
\begin{equation}
\left[ {\widetilde k}^2 - \left( \frac{\partial}{\partial z} - i A
t k_x\right)^2\right] u_z(k_x,k_y,z,t) = F(k_x,k_y,z),
\label{eq:app1}
\end{equation}
where $F(k_x,k_y,z)$ is the constant (in time) of integration and
defines the internal vortex mode perturbation in the flow:
$F(k_x,k_y,z)=0$ relates to a case when the internal perturbation
is absent.

Fourier transformation with respect to $z$:
\begin{equation}
\left[ \begin{array}{cc} u_z(k_x,k_y,z,t) \\
F(k_x,k_y,z)\end{array}\right] =
\frac{1}{2\pi}\int_{-\infty}^\infty
\left[\begin{array}{cc} u_z(k_x,k_y,k_z,t) \\
\widetilde F(k_x,k_y,k_z)\end{array}\right] e^{ik_zz}dk_z
\label{eq:fou1}
\end{equation}
reduces Eq.~(\ref{eq:app1}) to the following one:
\begin{equation}
k^2(t)
u_z(k_x,k_y,k_z,t)=\widetilde{F}(k_x,k_y,k_z)+4i\pi\widetilde k
C(k_x,k_y,t),\label{eq:app2}
\end{equation}
where
\begin{equation}
C\equiv\frac{1}{4i\pi\widetilde k}\left[ \left(\frac{d}{dz}
-2iAtk_x-ik_z\right)u_z(k_x,k_y,z,t)\right]\bigg|_{z=0}.\label{eq:app3}
\end{equation}

Defining $u_z(k_x,k_y,k_z,t)$ from Eq.~(\ref{eq:app2}), making
invert Fourier transform with respect to $k_z$, taking into
account boundary condition $|u_z|<\infty$ at $z=-\infty$ and the
fact that $C(k_z,k_y,t)$ does not depend on $z$, we obtain:
\begin{eqnarray}
&&u_z(k_x,k_y,z,t)=\frac{1}{2\pi}\int_{-\infty}^\infty
\frac{\widetilde{F}(k_x,k_y,k_z)}{k^2(t)}\exp(ik_zz)dk_z+ \nonumber \\
&&+ C(k_x,k_y,t)\exp[(\widetilde k + i A t k_x) z],
\label{eq:app4}
\end{eqnarray}
where $k^2(t)=\widetilde k^2 + k_z^2(t); ~k_z(t)\equiv k_z-Atk_x$.

The first term in Eq.~(\ref{eq:app4}) relates to the vortex mode
perturbation~\cite{CD90,C02}, whereas the second term, which is
exponentially decreasing with the depth, relates to the SFHs of
shear modified surface waves.

Substituting Eq.~(\ref{eq:app4}) into Eq.~(\ref{eq:pr1}), and
using boundary condition, Eq.~(\ref{eq:b2}), we obtain:
\begin{equation}
 \frac{d^2C}{dt^2}+ \frac{i A k_x}{\widetilde k}\frac{dC}{dt}+ \widetilde k g C =
 I(k_x,k_y,t),\label{eq:pr3}
\end{equation}
where
\begin{equation}
I(k_x,k_y,t)\equiv \int_{-\infty}^\infty \left[ 8 i A^2 k_x^2
\widetilde k \frac{k_z(t)}{k^6(t)}- \frac{\widetilde k
g}{k^2(t)}\right] \widetilde F(\b k)d k_z.\label{eq:pr4}
\end{equation}

Generally, Eqs.~(\ref{eq:pr3})-(\ref{eq:pr4}) describe the
dynamics of surface wave SFHs in the presence of the internal
vortical source - the term $I(k_x,k_y,t)$ is the result of an
interplay of the mean flow shear and the internal vortical
perturbations and couples the later perturbation with the surface
one. So, there is no coupling between these perturbations in the
absence of the shear. Indeed, if there are no surface
perturbations initially [$u_z(k_x,k_y,z=0,t=0)=0$], then from
Eqs.~(\ref{eq:app2}) and (\ref{eq:pr4}), one readily obtains that
$I(k_x,k_y,t) \sim u_z(k_x,k_y,z=0,t=0)$ at $A=0$, i.e.,
$I(k_x,k_y,t)\equiv0$. Thus, if there is no the source in
shearless flow initially, it does not appear afterward.

\section{SGWs and their generation in shear flow}

One can see from Eqs.~(\ref{eq:pr3}) and (\ref{eq:pr4}), that
there are two main effects of the shear: firstly, the second term
on the left hand side of Eq.~(\ref{eq:pr3}) indicates that the
velocity shear affects the frequencies of SGWs. Secondly, the
source term $I(k_x,k_y,t)$ caused by the internal perturbations,
couples the internal and surface perturbations and results the
emergence/generation of SGW in the flow. Our further attempts are
focused on the study of these effects.

\subsection{Shear modified SGWs}

In this subsection we study shear induced modifications of
properties of SGWs. For this purposes we assume that initially
there were no vortex mode perturbations: $\widetilde
F(k_x,k_y,k_z)=0$. Consequently, $I(k_x,k_y,t)=0$ [see
Eqs.~(\ref{eq:pr4})], and Eq.~(\ref{eq:pr3}) reduces to the
homogeneous one, with the solution:
\begin{equation}
C_h(k_x,k_y,t) = C_1(k_x,k_y)\exp(-i\Omega_1 t)+C_2(k_x,k_y)
\exp(-i \Omega_2t),\label{eq:so2}
\end{equation}
where $C_{1,2}(k_x,k_y)$ are determined by initial conditions and
\begin{equation}
\Omega_{1,2} =\pm\sqrt{ {\widetilde k g} +\frac{A^2
k_x^2}{4\widetilde k^2}}-\frac{A k_x}{2 \widetilde
k}=\sqrt{\widetilde k g} \left(\pm \sqrt{
{1+S^2}\frac{k_x^2}{\widetilde k^2}} -S\frac{k_x}{\widetilde k}
\right)\label{eq:so3}
\end{equation}
represents shear modified frequencies of SFH of SGW propagating in
opposite directions and $S\equiv A/(4\widetilde k g)^{1/2}$ is the
dimensionless shear rate. This equation provides, that in contrast
with acoustic and magnetohydrodynamic wave
modes~\cite{CTBM97,CCLT97,RPM00}, the presence of the shear does not
lead to the time variability of the frequency. However, velocity
shear leads to the non-trivial modification of the frequencies and
consequently phase velocities of SFH \cite{C68,S93}. Indeed, for the
value of the phase velocity, Eq.~(\ref{eq:so3}) provides:
\begin{equation}
V_{ph}(S,\phi)=\sqrt{\frac{g}{\widetilde
k}}\left(\sqrt{1+S^2\cos^2\phi}-S \cos\phi\right),\label{eq:va01}
\end{equation}
where $\phi\equiv\arccos(k_x/\widetilde k )$.

The phase velocity is isotropic in the shearless limit ($S = 0$),
while depends on $\phi$ in the shear flow. The anisotropy
increases with the shear rate. The value of the phase velocity is
minimal at $\phi=0$: $V_{ph}^{\rm min}=\sqrt{g/\widetilde
k}(\sqrt{1+S^2}-S)$ and is maximal at $\phi=\pi$: $V_{ph}^{\rm
max}=\sqrt{g/\widetilde k}( \sqrt{1+S^2}+S)$. Suppose SGW is
emitted by point source situated on the surface at $x=y=0$. From
Eq.~(\ref{eq:va01}) it follows that the propagation of the leading
wave crest is described by:
\begin{eqnarray}
r(S,\phi,t)&=&V_{ph}(S,\phi)t= \nonumber \\
&=&\sqrt{\frac{g}{\tilde k}}\left(\sqrt{1+S^2\cos^2\phi}-S
\cos\phi\right)t. \label{eq:va02}
\end{eqnarray}
Figure \ref{fig:fig1} shows the leading wave crest of the SGW for
three different moments of time $t_1,~t_2,~t_3$, when $t_2=2t_1,~
t_3=3t_1$, that are circular but not concentric.

\subsection{Generation of SGWs by internal vortices}

Let us initially analyze the source term $I(k_x,k_y,t)$ that is
determined by $\widetilde F(k_x,k_y,k_z)$. Assume $\widetilde
F(k_x,k_y,k_z)$ is a localized function in the wavenumber space,
with a center of localization at $\b k_0 =(k_{x0},k_{y0},k_{z0})$.
Note that the first multiplier in the integrand of
Eq.~(\ref{eq:pr4}) reaches its maximum when $k_z-Ak_xt=0$.
Consequently, the maximum of the integral is in the vicinity of
time $t=t_\ast \equiv k_{z0}/(Ak_{x0})$. Equation~(\ref{eq:pr4})
provides, that generally $I(k_x,k_y,t)$ tends to zero in both
limits $t\rightarrow\pm\infty$. Actually, it exists some time
interval $2\Delta t$ around $t_\ast$ where the source term differs
from zero. The value of $\Delta t$ depends on the degree of
localization of internal perturbation, i.e., of $\widetilde
F(k_x,k_y,k_z)$ in wavenumber space. (The source localization is
demonstrated below on a specific example.) Thus, in the case of a
localize source, the coupling between surface (gravity wave) and
internal (vortex mode) perturbations takes place in some time
interval $2\Delta t$ around $t_\ast$, and at $|t-t_\ast|>\Delta t$
this perturbations can be considered separately.

The general solution of the inhomogeneous equation,
Eq.~(\ref{eq:pr3}), is the sum of the general solution of
corresponding homogenous equation and a partial solution of the
equation:
\begin{equation}
C(k_x,k_y,t)=C_h(k_x,k_y,t)+C_i(k_x,k_y,t).\label{eq:c1}
\end{equation}
The general solution $C_h(k_x,k_y,t)$ is presented by
Eq.~(\ref{eq:so2}), whereas a partial solution of Eq.
(\ref{eq:pr3}) is:
\begin{eqnarray}
C_i&=&\frac{1}{2\Omega_0}\exp{(-i\Omega_1 t)}\int^{t}_{t_0}
I(k_x,k_y,t^\prime) {\exp{(i\Omega_1 t^\prime)}} d t^\prime\nonumber\\
& &\mbox{} -\frac{1}{2\Omega_0} \exp{(-i\Omega_2 t)}\int^{t}_{t_0}
I(k_x,k_y,t^\prime) \exp{(i\Omega_2 t^\prime)} d
t^\prime,\label{eq:so4}
\end{eqnarray}
where
\begin{equation}
\Omega_{0} = \sqrt{ {\widetilde k g} +\frac{A^2 k_x^2}{4\widetilde
k^2}} = \sqrt{\widetilde k g} \sqrt{1 +S^2\frac{k_x^2}{\widetilde
k^2}}. \label{eq:va03}
\end{equation}

Assume that the coupling between the surface and internal modes
can be neglected at the initial moment of time $t_0$, i.e.,
$t_0<t_\ast-\Delta t$. After passing through the coupling time
interval, for any $t>t_f=t_\ast+\Delta t$ the modes become
independent again. However, during the time interval $[t_0,t_f]$,
internal vortices generate SGWs with frequencies $\Omega_1$ and
$\Omega_2$ [see Eq. (\ref{eq:so3})]. As it follows from
Eqs.~(\ref{eq:so2}), (\ref{eq:c1}), and (\ref{eq:so4}), if
initially there are no SGWs ($C_{1,2}=0$), the generated SFH
amplitudes ($Q_{1,2}$) are:
\begin{equation}
Q_1(k_x,k_y)= \frac{1}{2\Omega_0}\left|\int^{t_f}_{t_0}
I(k_x,k_y,t^\prime) \exp{(i\Omega_1 t^\prime)} d t^\prime \right
|,\label{eq:am1}
\end{equation}
\begin{equation}
Q_2(k_x,k_y)= \frac{1}{2\Omega_0}\left | \int^{t_f}_{t_0}
I(k_x,k_y,t^\prime) \exp{(i\Omega_2 t^\prime)} d
t^\prime\right|.\label{eq:am2}
\end{equation}
One can replace the limits of integration by $\pm \infty$. After
integration in time this yields:
\begin{eqnarray}
Q_{1,2}&=&\frac{\pi k_x}{\widetilde k^3}
\left(\frac{A}{2\Omega_0}\mp \frac{k_x}{\widetilde
k}\right)\exp\left[-
\frac{(\Omega_{0}\mp A/2)\widetilde k}{A k_x}\right]\nonumber \\
& &\mbox{}\times\int\widetilde F(k_x,k_y,k_z) \exp\left(-i
\frac{(\Omega_{0}\mp A/2)k_z}{A k_x}\right) d k_z\nonumber \\
&=&\frac{2\pi^2 k_x}{\widetilde k^3} \left( \frac{A}{2\Omega_0}\mp
\frac{k_x}{\widetilde k}\right ) \exp\left[-
\frac{(\Omega_{0}\mp A/2)\widetilde k}{A k_x}\right]\nonumber \\
& &\mbox{}\times F\left(k_x,k_y,- \frac{\Omega_{0}\mp A/2}{A
k_x}\right), \label{eq:am3}
\end{eqnarray}

Note that the last multipliers in Eq.~(\ref{eq:am3}) are
proportional to the vorticity of initial perturbations at $z_{1,2}=-
(\Omega_{0}\mp A/2)/(A k_x)$ respectively. The second multipliers
indicates, that at small shear rates ($S \equiv A/ \sqrt{4\widetilde
k g} \ll 1$), the amplitudes of generated SGWs are exponentially
small with respect to the large parameter $1/S$.
Equation~(\ref{eq:am3}) also indicates that for a fixed $k_x$, the
generation has maximal effectiveness in two dimensional case
($k_y=0$).

Let us describe the dynamic picture for a specific example, when
pure internal vortex mode perturbation (without any admix of
surface waves) is imposed in the flow initially. For simplicity we
consider {\it two dimensional} problem, when $\partial/\partial y
= 0$. The vertical velocity of the imposed perturbation is given
by:
\begin{eqnarray}
u_z(x,z,t_0)&=&z^3\eta(-z)\exp\left(
-\frac{[(z+z_0)\cos\phi+x\sin\phi]^2}{L_1^2} \right)\nonumber \\
&&\mbox{}\times\exp\left(-\frac{[(z+z_0)\sin\phi-x\cos\phi]^2}{L_2^2}\right),
\label{eq:sp1}
\end{eqnarray}
where $\eta(z)$ is Heaviside function, $(0,-z_0)$ is the center of
the localization, $L_{1,2}$ characterize vertical and horizontal
scales respectively and $\phi$ is the slope of the perturbation.

Numerical solution of the problem was performed as follows:
Fourier transform of Eq.~(\ref{eq:sp1}) with respect to $x$
variable allows to determine $F(k_x,z)$ through
Eq.~(\ref{eq:app1}). Another Fourier transform with respect to $z$
yields $\widetilde F(k_x,k_z)$. Then, the source function
$I(k_x,t)$ is found by Eq.~(\ref{eq:pr4}). Thus, the solution of
the problem for a fixed $k_x$ reduces to the numerical solution of
the inhomogeneous equation, Eq.~(\ref{eq:pr3}), with known
$I(k_x,t)$.

Dependence of the source function $I(k_x,t)$ on $t$ at $L_1=1$,
$L_2=7$, $\phi=\pi/18$, $k_x=1$, $z_0=2$ for two different values of
the shear rate $S=0.08$ (dashed line) and $S=0.32$ (solid line) is
presented in Fig.~\ref{fig:fig2}. As it was mentioned above, the
source term is localized function and considerably differs from zero
only in the interval $t\in (20,40)$ for $S=0.08$ and $t\in (5,10)$
for $S=0.32$.

For analysis of the wave generation effectiveness it is useful to
introduce the generation coefficients, that characterize the ratio
of generated wave energy density and the maximum energy density of
the initial vortex mode perturbations for a fixed value of $k_x$.
Taking into account that the maximum energy density of the vortex
mode perturbations is:
\begin{equation}
E_v = \frac{1}{2k_x^4} \int_{-\infty}^\infty |F(k_x,z)|^2
dz,\label{eq:e1}
\end{equation}
and the energy density of the generated waves:
\begin{equation}
E_{w1,2} = \frac{1}{k_x} Q_{1,2}^2(k_x),\label{eq:e2}
\end{equation}
we define the non-dimensional generation coefficients as:
\begin{equation}
G_{1,2} = Q_{1,2}(k_x)\left(\frac{2k_x^3}{\int_{-\infty}^\infty
|F(k_x,z)|^2 dz}\right)^{1/2} .\label{eq:e3}
\end{equation}

Figure~\ref{fig:fig3} represents generation coefficients $G_1$
(dashed line) and $G_2$ (solid line) vs shear rate $S$ at $L_1=1$,
$L_2=7$, $\phi=\pi/18$, $k_x=1$ and $z_0=2$. As it can be seen, at
small values of the shear rate, generation of SGW with frequency
$\Omega_1$ (i.e., propagating along $x$ axis) considerably exceeds
the generation of SGW with frequency $\Omega_2$ (i.e., propagating
against $x$ axis), whereas the later wave is generated more
effectively at $S>0.15$.

The wave generation is well traced in Figs.~\ref{fig:fig4} and
\ref{fig:fig5}, where the temporal evolution of the vertical
component of velocity perturbation at the surface obtained by
numerical solution of Eqs.~(\ref{eq:pr3})-(\ref{eq:pr4}) is
presented for $S=0.32$ and $S=0.08$ respectively. The other
parameters are the same as in Fig.~\ref{fig:fig2}. Purely internal
vortex mode perturbation is imposed in the equations initially.
The generation takes place in the time interval where $I(k_z,t)$
noticeably differs from zero. Afterward, just (two) waves with
different frequencies and amplitudes exist. At $S=0.32$, presented
in Fig.~\ref{fig:fig4}, the generation takes place in the time
interval $t\in (5,10)$. Besides, SGW propagating against $x$ axis
is mainly generated. In contrast to this, at $S=0.08$, presented
in Fig.~\ref{fig:fig5}, the generation of SGW propagating along
$x$ axis is dominated. These numerical results are in agreement to
analytical ones [see Eq.~(\ref{eq:am3}) and Fig.~\ref{fig:fig3}].

\section{Discussion}

In the previous sections simplified model was considered that
allowed us to simplify mathematical description and study shear
induced effects in the 'pure' form. For instance, the influence of
the viscosity was ignored and the density ratio $\rho_a/\rho_0$ of
the fluids above and below of the surface $z=0$ was assumed to be
zero. The later assumption allows to ignore all the dynamical
processes in the upper fluid. On the other hand, it is well known
that in the case of ocean waves the wind is the most important and
powerful source of the waves. In this section we discuss possible
applications of the studied linear effects to the concrete physical
applications.

\subsection{Ocean waves}

It is well known \cite{P57,M57,M59,M61,M62} that the wind is main
source of ocean SGWs. In the context of future discussion the papers
of Chalikov group \cite{C76,CM91} should also be noted, where the
influence of small scale turbulence in the air on wave growth was
studied in detail. At present there exists well developed theory
both SGW generation and nonlinear evolution that is mainly confirmed
by experiments as well as numerical simulations (for recent review
see, e.g., \cite{J}). After development of wind driven instability
nonlinear 4-wave resonant interactions transfer the wave energy to
smaller scales. Existing theory predicts that for relatively small
frequencies Zakharov-Philonenko \cite{ZP67} spectrum $E(\omega)\sim
\omega^{-4}$ of SGW fluctuations (sometimes called Toba's spectrum)
should be observed (in this context see also \cite{K83}), whereas
for relatively high wave numbers nonlinearity becomes strong and
Phillips's spectrum $E(\omega)\sim \omega^{-5}$ of the wave
turbulence should develop. Existing observations confirm this
predictions and provide that in the range $\omega_p/3 < \omega <3
\omega_p$, where $\omega_p$ is the peak frequency,
Zakharov-Philonenko spectrum is usually observed. For $\omega > 3
\omega_p$ the spectrum becomes very close to Phillips's one
\cite{J}. The properties of the wave spectrum in the very short
wavelength region, as well as dynamics of SGW turbulent fluctuations
dissipation is much more unclear \cite{Z99}.

In the case of ocean waves presented linear mechanism of SGW
generation can have important contribution to the balance of small
scale SGW fluctuations. Indeed, characteristic length scale of the
turbulence at the ocean surface is much smaller than in the air.
Namely, the characteristic length and velocity scales are $u_\ast
\sim 1~cm/sec$ and $l \sim 1~cm$ respectively \cite{KL83}. On the
other hand, in the presence of the wind strong velocity shear $A
\sim 10~sec^{-1}$ is presented in so called 'buffer layer'
\cite{CPB99} of the water, with the thickness $l_1 \sim
(20-100)l_0$, where $l_0 \approx \nu/u_\ast$ is the dissipation
length scale and $\nu$ is kinematic viscosity of the water.  Simple
estimates yields $l_1 \sim (0.5-1)~cm$. Presented linear mechanism
implies that vortical perturbations generate SGWs with the same
length scale. Therefore in the case of ocean waves internal vortex
mode perturbations should effectively generate small scale SGWs -
with the wavelength just above the capillary length scale
$\lambda_c=0.39~cm$ \cite{LL}. In this context the study of the
influence of capillary effects on the processes discussed above
seems to be interesting. Analysis of this problem will be presented
elsewhere.

\subsection{Interfacial gravity waves}

In the analysis presented in Secs. II and III density ratio
$\rho_a/\rho_0$ of the fluids above and below of the surface $z=0$
was assumed to be zero. Obtained results can be readily generalized
in the case of interfacial GWs. If the density of upper and lower
fluids are $\rho_1$ and $\rho_2$ and the shear rates are $A_1$ and
$A_2$ respectively, then shear modified dispersion of interfacial
GWs is given by the same expression (\ref{eq:so3}) with $g$ and $A$
replaced by $g_\ast$ and $A_\ast$, where
\begin{equation}
g_\ast=g\frac{\rho_2-\rho_1}{\rho_2+\rho_1},~~~A_\ast=\frac{A_2\rho_2-A_1\rho_1}{\rho_2+\rho_1}.
\end{equation}
From this equation it follows, that the influence of shear both on
the wave dispersion and coupling with internal vortex perturbations,
that is determined by dimensionless parameter
\begin{equation}
S_\ast \equiv \frac{A_\ast}{\sqrt{4\tilde k g_\ast}}=
S_2\frac{1-\rho_1 A_1/\rho_2 A_2}{\sqrt{1-\rho_1^2/\rho_2^2}}
\end{equation}
is much more notable when the fluids have comparable densities if
$\rho_1A_1$ is not very close to $\rho_2A_2$. Therefore, described
shear induced effects usually should have much more influence on the
dynamics of interfacial waves, then on the ocean waves.

\section{Summary}

\begin{figure}
\includegraphics{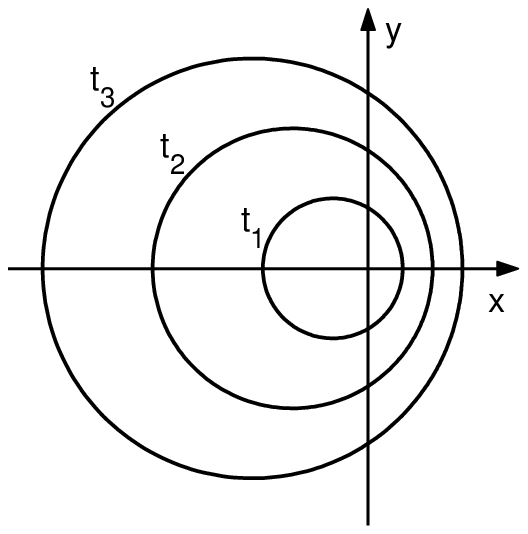}
\caption{\label{fig:fig1} }
\end{figure}

\begin{figure}
\includegraphics[]{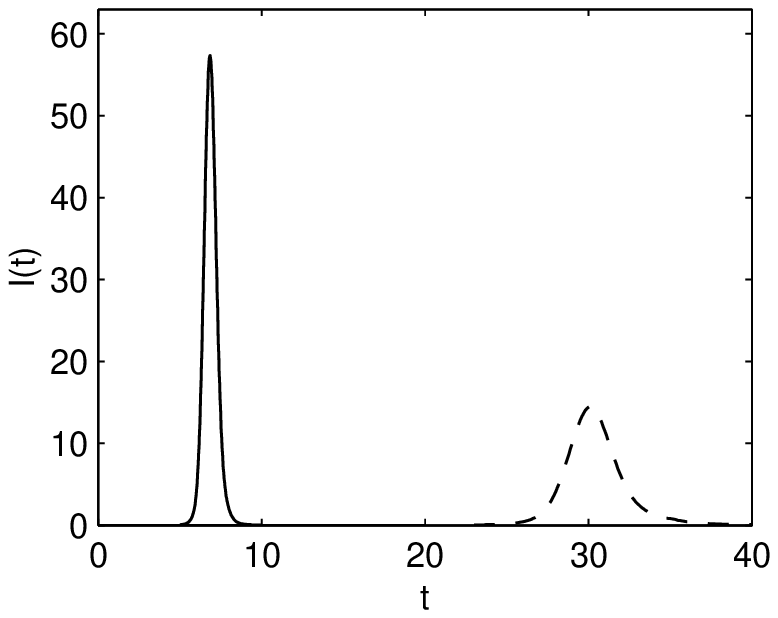}
\caption{\label{fig:fig2}}
\end{figure}

\begin{figure}
\includegraphics[]{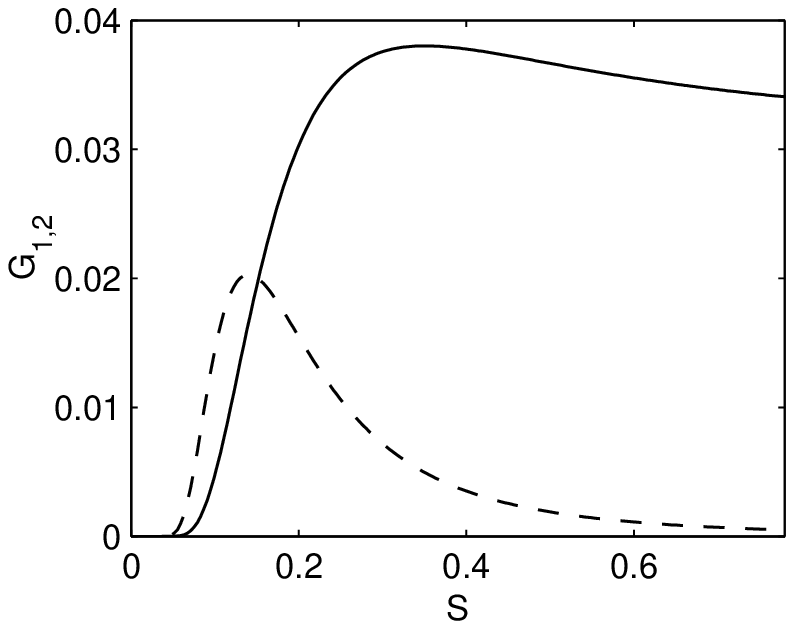}
\caption{\label{fig:fig3} }
\end{figure}

\begin{figure}
\includegraphics[]{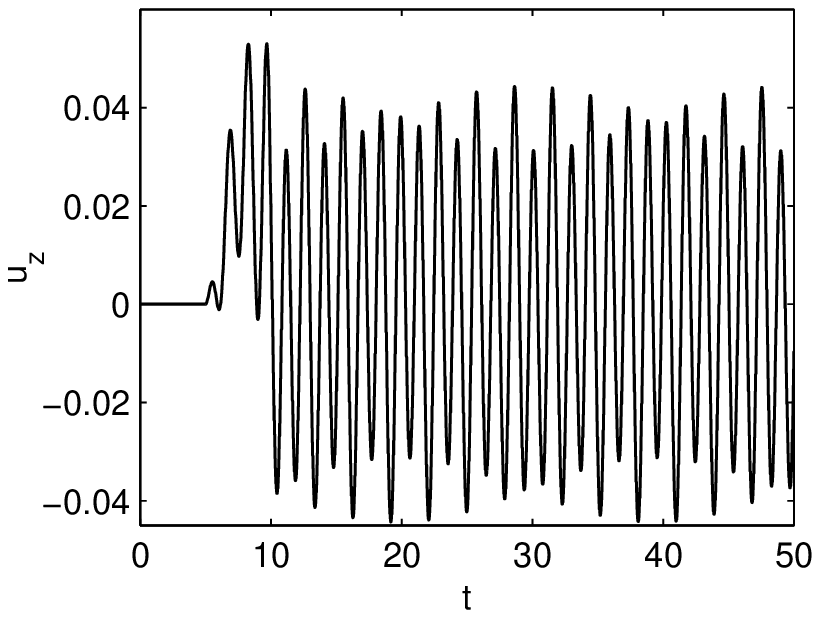}
\caption{\label{fig:fig4} }
\end{figure}

\begin{figure}
\includegraphics[]{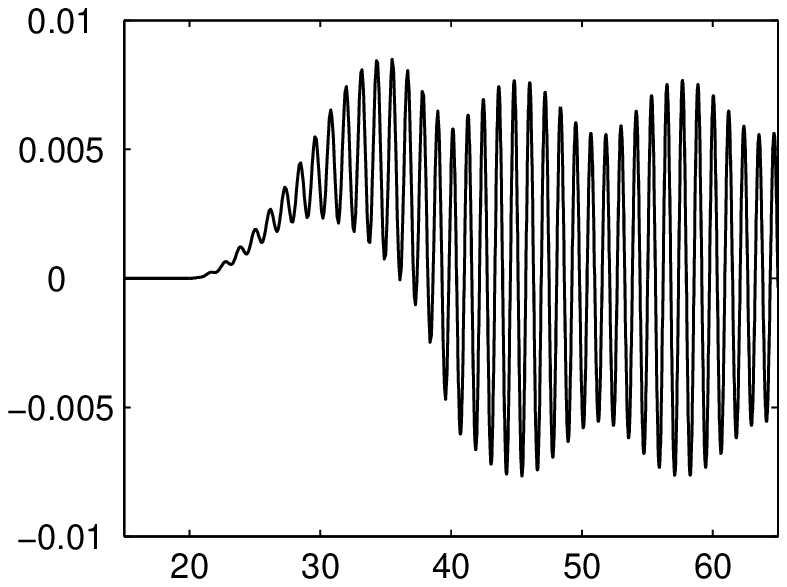}
\caption{\label{fig:fig5} }
\end{figure}

Let us summarize the main features of the linear dynamics of
surface gravity weaves in a simplified deep fluid (at $z< 0$) flow
with vertical shear of the mean velocity $\b U_0=(A z,0,0)$. The
simplification lies in the neglecting of the fluid compressibility
and stratification, in other words, in the consideration of the
system containing just two modes of perturbation: surface gravity
wave mode and internal vortex mode. Special features of SGW in the
system are the following:

The mean flow velocity shear causes the non-trivial modification
of the frequencies and phase velocities of SGWs. The frequencies
are defined by Eq.~(\ref{eq:so3}). The phase velocity becomes
anisotropic [see Eq.~(\ref{eq:va01}) and Fig.~\ref{fig:fig1}]: its
value is minimal for SFH propagating along $x$ axis [$V_{ph}^{\rm
min}=\sqrt{g/\widetilde k}(\sqrt{1+S^2}-S)$] and maximal for SFH
propagating against $x$ axis [$V_{ph}^{\rm min}=\sqrt{g/\widetilde
k}(\sqrt{1+S^2}+S)$].

The mean flow velocity shear leads to the appearance of the
intrinsic (to the fluid) source of SGW generation via coupling the
wave with the internal vortex mode perturbations - the coupling
results the emergency/generation of SGWs by internal vortex mode
perturbations at $S \gtrsim 0.05$. The generation is different for
the waves propagating in the different directions [see
Eq.~\ref{eq:am3}]. Generation of SGW with frequency $\Omega_1$
considerably exceeds the generation of SGW with frequency
$\Omega_2$ for relatively small shear rates $S$, whereas the later
wave is generated more effectively for the high shear rates
($S>0.15$).

\begin{acknowledgments}

This research is supported by ISTC grant G 553.

The work was supported in part by the Department of Energy Grant
No.~DE-FG03-96ER-54346.

\end{acknowledgments}

\newpage

FIG. 1: Shear induced anisotropy of SGW propagation: The leading
wave crest at three different moments of time $t_1,~t_2,~t_3$,
when $t_2=2t_1,~ t_3=3t_1$,  that are circular, but not
concentric. A point source of the SGW is located at $x=y=0$.

FIG. 2: $I(k_x,k_y,t)$ vs time at $S=0.32$ (solid line) and $S=0.08$
(dashed line) $~k_x=1,~L_1=1,~L_2=7,~z_0=2$ and $\phi=\pi/18$.

FIG. 3: Generation coefficients $G_1$ (dashed line) and $G_2$ (solid
line)  vs. shear rate $S$ at $k_x=1,~L_1=1,~L_2=7,~z_0=2$ and
$\phi=\pi/18$.

FIG. 4: $u_z(k_x,t)$ vs. time at
$S=0.32,~k_x=1,~L_1=1,~L_2=7,~z_0=2$ and $\phi=\pi/18$.

FIG. 5: $u_z(k_x,t)$ vs. time at
$S=0.08,~k_x=1,~L_1=1,~L_2=7,~z_0=2$ and $\phi=\pi/18$.

\newpage


\end{document}